\documentclass[review]{elsarticle}
\journal{Nucl. Instrum. Meth. A}
\usepackage{subfigure}
\usepackage{graphicx}
\usepackage{xcolor}
\usepackage{placeins}

\usepackage{amsmath}
\usepackage{bm}  

\begin{document}

\begin{frontmatter}

\title{Design and commissioning of a windowless gas-target system for high-current beams at JUNA}

\author[1]{Xuyang Wang\fnref{equal}}
\ead{12432067@mail.sustech.edu.cn}

\author[2,3]{Fuqiang Cao\fnref{equal}}

\author[1]{Weike Nan\corref{cor1}}
\ead{nanwk@sustech.edu.cn}

\author[2,3]{Gang Lian\corref{cor1}}
\ead{lgang@ciae.ac.cn}

\author[2,3]{Wei Nan}

\author[2]{Hongwei Huang}

\author[2,4,5]{Yuchen Jiang}

\author[2]{Jingyu Dong}

\author[2]{Minghao Zhu}

\author[2]{Yuwen Chen}

\author[2]{Yihui Liu}

\author[2,6]{Jinlong Ma}

\author[1]{Yueran Wang}

\author[1]{Jie Chen}

\author[2,3]{Yunju Li}

\author[2,3]{Shengquan Yan}

\author[2,3]{Youbao Wang}

\author[7]{Xiao Fang}

\author[8]{Luohuan Wang}

\author[2,3]{Yangping Shen}

\author[2,3,9]{Bing Guo}

\author[1,2,3]{Weiping Liu\corref{cor1}}
\ead{wpliu@ciae.ac.cn}

\cortext[cor1]{Corresponding authors.}
\fntext[equal]{These authors contributed equally to this work.}

\affiliation[1]{
    organization={Department of Physics, Southern University of Science and Technology},
    city={Shenzhen},
    postcode={518055},
    state={Guangdong},
    country={China}
}

\affiliation[2]{
    organization={China Institute of Atomic Energy},
    city={Beijing},
    postcode={102413},
    country={China}
}

\affiliation[3]{
    organization={Jinping Deep Underground Frontier Science and Dark Matter Key Laboratory of Sichuan Province},
    city={Liangshan},
    postcode={615000},
    country={China}
}

\affiliation[4]{
    organization={Institute of High Energy Physics, Chinese Academy of Sciences},
    city={Beijing},
    postcode={100049},
    country={China}
}

\affiliation[5]{
    organization={University of Chinese Academy of Sciences},
    city={Beijing},
    postcode={100049},
    country={China}
}

\affiliation[6]{
    organization={College of Nuclear Science and Technology, Beijing Normal University},
    city={Beijing},
    postcode={100875},
    country={China}
}

\affiliation[7]{
    organization={Sino-French Institute of Nuclear Engineering and Technology, Sun Yat-sen University},
    city={Zhuhai},
    postcode={519082},
    country={China}
}

\affiliation[8]{
    organization={School of Mathematics and Physics, Handan University},
    city={Handan},
    postcode={056005},
    country={China}
}

\affiliation[9]{
    organization={School of Physics, Xi'an Jiaotong University},
    city={Xi'an},
    postcode={710049},
    country={China}
}

\begin{abstract}
Windowless gas targets avoid the beam-energy loss and straggling introduced by entrance foils and are therefore well suited for direct measurements of low-energy nuclear reactions. A windowless gas-target system designed for operation with milliampere beams has been developed for the Jinping Underground Nuclear Astrophysics facility (JUNA). The system combines three-stage differential pumping, closed-loop gas recovery and purification, a constant-temperature power-compensation calorimeter, and a position-resolved target-thickness monitor based on secondary elastic scattering. Stable operation was achieved over a target-pressure range of \(1\text{--}3\,\mathrm{mbar}\), with pressure fluctuations below \(1\%\) during \(8\,\mathrm{h}\) of continuous circulation, while the accelerator-side pressure was maintained at approximately \(10^{-4}\,\mathrm{Pa}\). The closed-loop gas-circulation system maintained stable target conditions, while gas-transport calculations indicated that the axial pressure nonuniformity remained within approximately \(1.6\%\) under representative operating conditions. Calorimeter measurements were consistent with the thermal calculations, supporting the sensitivity correction used for beam-power determination. Beam commissioning with \(^{14}\mathrm{N}(p,\gamma)^{15}\mathrm{O}\) and \(^{12}\mathrm{C}(p,\gamma)^{13}\mathrm{N}\) at the \(600\,\mathrm{kV}\) Cockcroft--Walton accelerator of the China Institute of Atomic Energy (CIAE) demonstrated stable operation of the gas-target and \(\gamma\)-ray detection systems and provided information on the influences of reaction position and beam heating. These results demonstrate the operating stability and diagnostic capability of the system for future high-current, low-energy nuclear-reaction measurements at JUNA.

\end{abstract}



\end{frontmatter}


\section{Introduction}
Nuclear astrophysics seeks to understand the origin and evolution of the elements through the nuclear reactions occurring in stars and other astrophysical environments. At the Gamow energies characteristic of stellar burning, charged-particle reaction cross sections are strongly suppressed by the Coulomb barrier, and direct measurements are limited by both extremely low reaction yields and environmental backgrounds~\cite{Adelberger2011SolarFusionII,Aliotta2022DirectMeasurements}. Underground laboratories substantially suppress cosmic-ray-induced backgrounds and thereby provide favorable conditions for extending direct measurements towards lower energies~\cite{Costantini2009LUNA,Szucs2012ShallowUnderground,Broggini2018LUNAStatus}. For nuclides such as hydrogen, helium, and neon that are commonly employed in gaseous form, gas targets provide an important experimental approach. Conventional gas cells confine the target gas with entrance foils, but the additional beam-energy loss and energy straggling introduced by the foil increase the uncertainty in the reaction energy and become particularly detrimental at low beam energies. Windowless gas targets avoid this limitation by using differential pumping to connect a finite-pressure target chamber directly to the high-vacuum accelerator beam line~\cite{Casella2002UndergroundSetup}.

Windowless gas targets have enabled a number of important measurements in nuclear astrophysics. At LUNA, differentially pumped recirculating gas targets have been used for direct measurements of the key pp-chain reaction \(^{3}\mathrm{He}(\alpha,\gamma)^{7}\mathrm{Be}\)~\cite{Confortola2007He3ag,Costantini2008He3ag}, the CNO-cycle bottleneck reaction \(^{14}\mathrm{N}(p,\gamma)^{15}\mathrm{O}\)~\cite{Costantini2009LUNA}, and \(^{22}\mathrm{Ne}(p,\gamma)^{23}\mathrm{Na}\)~\cite{Cavanna2014LUNAGasTarget,Ferraro2018LUNAGasTarget}. In particular, the windowless-gas-target measurement of \(^{14}\mathrm{N}(p,\gamma)^{15}\mathrm{O}\) extended the direct determination of the total cross section towards stellar energies and substantially improved the experimental constraint on the CNO-cycle reaction rate. In addition to extended static gas targets, facilities such as JENSA have developed high-density windowless gas-jet targets for reaction studies in inverse kinematics and with rare-isotope beams~\cite{Schmidt2018JENSA}. These developments demonstrate that windowless gas targets can provide sufficient target thickness while eliminating the entrance-foil contribution to the beam-energy uncertainty.

At beam currents approaching the milliampere level, however, the main challenges shift from the entrance window to target operation and beam-on diagnostics. The increased gas load must be handled while maintaining both the target pressure and the high vacuum required by the accelerator, and enriched target gas must be efficiently recovered and purified. Beam heating reduces the local gas density along the beam path, causing the effective target thickness to differ from that inferred from static pressure and temperature measurements~\cite{Gorres1980BeamHeating}. In addition, charge exchange between the beam and target gas modifies the transmitted charge-state distribution and can compromise beam normalization based on Faraday-cup charge integration~\cite{Allison1958ChargeChanging,Meckbach1967}. High-current windowless gas targets therefore require not only stable differential pumping and gas circulation, but also independent diagnostics of the incident-particle number and effective target thickness under beam-on conditions.

The Jinping Underground Nuclear Astrophysics facility (JUNA) was established at the China Jinping Underground Laboratory to perform direct low-energy nuclear-reaction measurements with intense light-ion beams~\cite{Liu2016JUNA,Liu2022JUNACommissioning,Kajino2023JUNA}. In this work, we present a windowless gas-target system developed for high-current operation at JUNA, incorporating three-stage differential pumping, closed-loop gas recovery and purification, a constant-temperature power-compensation calorimeter, and a position-resolved target-thickness monitor based on secondary elastic scattering. Off-beam circulation tests and gas-transport simulations are used to characterize the pressure stability and static axial pressure distribution in the target chamber under closed-loop operation. Experimental measurements and thermal simulations are used to characterize the calorimeter response and determine the sensitivity correction required for beam-power measurement. Beam commissioning with \(^{14}\mathrm{N}(p,\gamma)^{15}\mathrm{O}\) and \(^{12}\mathrm{C}(p,\gamma)^{13}\mathrm{N}\) was performed using the \(600\,\mathrm{kV}\) Cockcroft--Walton accelerator at the China Institute of Atomic Energy (CIAE) to evaluate the combined operation of the windowless gas target and the \(\gamma\)-ray detection system and to investigate the target response under beam irradiation.

\section{Windowless gas-target design and diagnostic development}
\begin{figure}[htbp]
    \centering
    \includegraphics[width=0.95\textwidth]{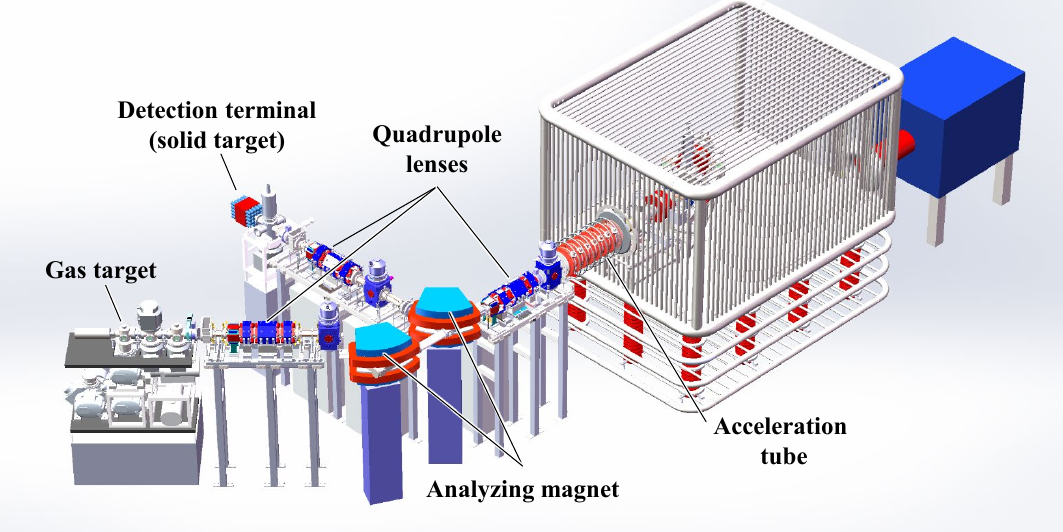}
    \caption{Schematic layout of the JUNA experimental terminal.}
    \label{fig:juna_terminal_layout}
\end{figure}

JUNA exploits the low-background environment of the Jinping underground laboratory and its high-intensity beams to perform direct measurements of key nuclear reactions relevant to stellar evolution and nucleosynthesis. Several reactions of interest, including the neutron-source reaction \(^{22}\mathrm{Ne}(\alpha,n)^{25}\mathrm{Mg}\) in massive stars and the solar-fusion reaction \(^{3}\mathrm{He}(\alpha,\gamma)^{7}\mathrm{Be}\), require isotopically enriched gaseous targets. Accordingly, in addition to the \(90^{\circ}\) solid-target station, JUNA is equipped with a windowless gas-target station located on the \(45^{\circ}\) beam line. The overall layout of the JUNA accelerator and experimental stations is shown in Fig.~\ref{fig:juna_terminal_layout}

At the low energies relevant to nuclear astrophysics, the small reaction cross sections impose stringent requirements on beam intensity, effective target thickness, target stability, and reaction-energy control. For example, LUNA employed a windowless \(^{3}\mathrm{He}\) gas target operated at approximately \(0.7,\mathrm{mbar}\), together with a \(^{4}\mathrm{He}^{+}\) beam of about \(250,\mu\mathrm{A}\), for direct measurements of the \(^{3}\mathrm{He}(\alpha,\gamma)^{7}\mathrm{Be}\) reaction~\cite{Costantini2008He3ag}. JUNA, in comparison, can provide helium beams at the milliampere level~\cite{Liu2025JUNAProgressNPA}. While the increased beam intensity improves the achievable reaction yield, it also places more stringent requirements on the operation and diagnostics of a windowless gas target. The larger gas load requires stable differential pumping and gas circulation, beam heating modifies the local gas density along the beam path and hence the effective target thickness, and charge exchange between the beam and target gas reduces the reliability of conventional Faraday-cup measurements. High-current operation therefore requires not only a stable pressure gradient between the target chamber and the accelerator beam line, but also dedicated diagnostics for the effective target thickness and incident-particle flux.

To address these requirements, the present system was developed in three areas: gas circulation, effective target-thickness monitoring, and incident-particle flux determination. It comprises a closed-loop gas-recovery and purification system, a position-resolved target-thickness monitor based on secondary elastic scattering, and a constant-temperature power-compensation calorimeter. The following sections describe the design and operating principles of these subsystems, together with the corresponding experimental tests and numerical simulations.

%


\subsection{Gas Circulation System}
\begin{figure}[htbp]
    \centering
    \includegraphics[width=0.80\textwidth]{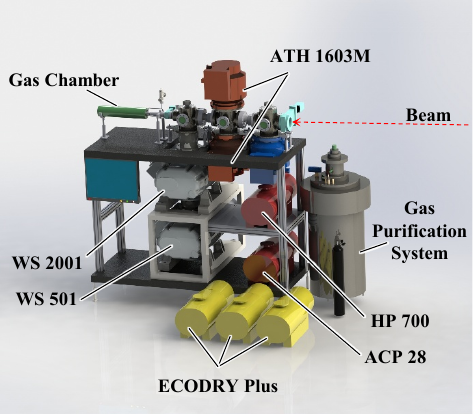}
    \caption{Windowless gas-target station.}
    \label{fig:windowless_target_photo}
\end{figure}

During operation of a windowless gas target, target gas continuously escapes through the conductance-limiting tubes into the differential-pumping stages and is removed by the corresponding pump trains. Direct discharge of the pump exhaust would result in substantial consumption of isotopically enriched gas. Recirculation is therefore necessary for long-duration operation, but repeated circulation may introduce or accumulate residual impurities from outgassing and small leaks. The present system consequently combines gas recovery with purification in a closed loop: the exhaust from the differential-pumping stages is collected, purified, and returned to the target chamber. The windowless gas-target station is shown in Fig.~\ref{fig:windowless_target_photo}.

\begin{figure}[htbp]
    \centering
    \includegraphics[width=0.95\textwidth]{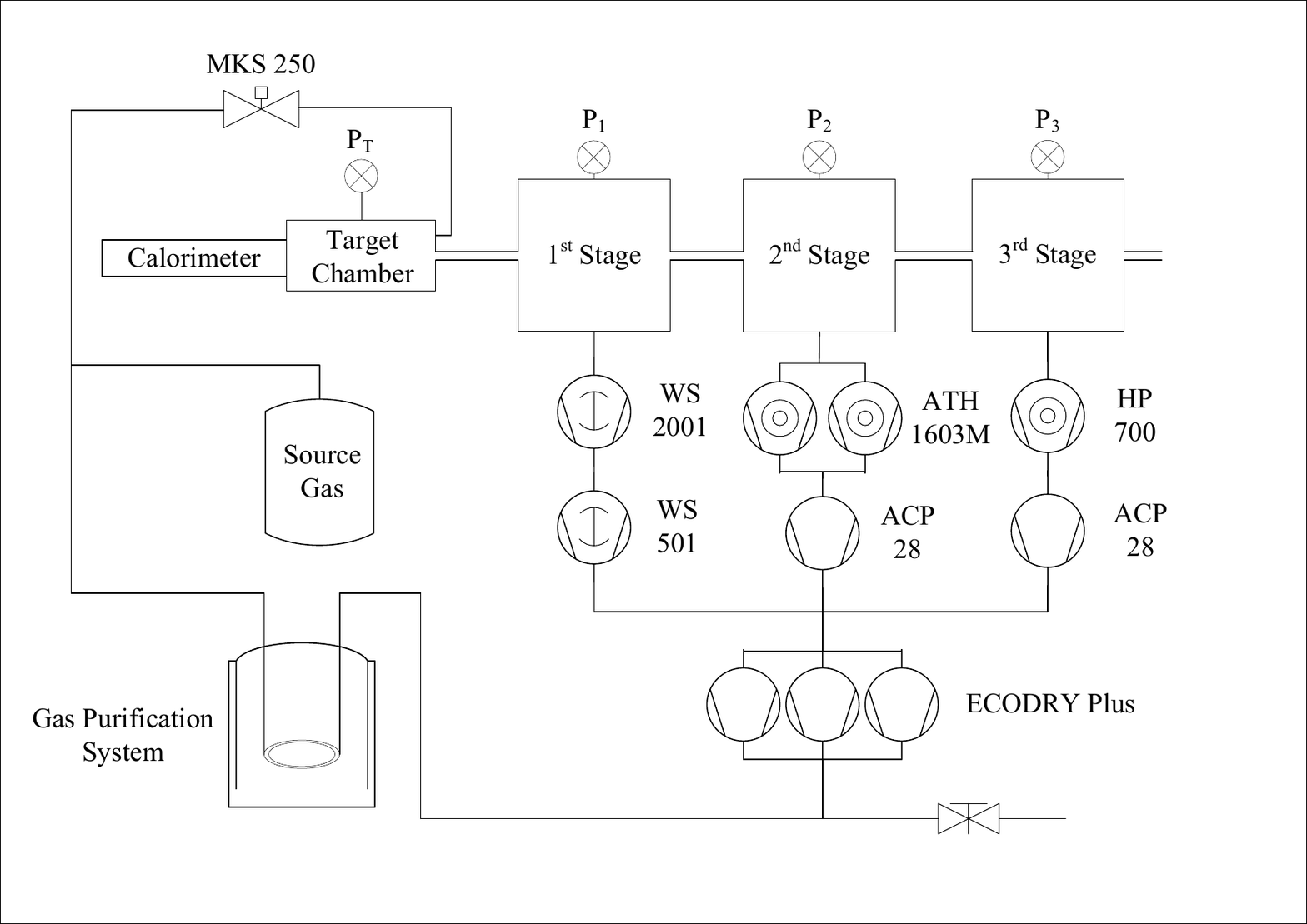}
    \caption{Schematic of the target chamber, three-stage differential-pumping system, and closed-loop gas-circulation and purification unit.}
    \label{fig:gas_circulation}
\end{figure}

A schematic of the target chamber, three differential-pumping stages, and gas-circulation and purification unit is shown in Fig.~\ref{fig:gas_circulation}. The first differential stage receives the largest gas load and is evacuated by a Roots-pump train. The second and third stages employ turbomolecular pumps backed by dry pumps to progressively reduce the pressure towards the accelerator beam line. The exhaust lines of the pump trains are combined into a common recovery line. The recovered gas passes through condensation and zeolite-adsorption units to remove condensable species and residual impurities accumulated during circulation. A make-up gas branch between the purification and gas-supply units compensates for unavoidable gas losses during operation and maintenance. The purified gas is then returned to the target chamber through a regulating valve controlled by an MKS~250 unit. In this way, differential pumping, gas recovery, purification, loss compensation, and controlled gas supply form a continuous closed-loop system.

Gas recovery and pressure control are also important features of other windowless gas-target systems used for nuclear astrophysics. The recirculating static gas target at LUNA has been operated over target pressures of approximately \(0.5\text{--}4.0\,\mathrm{mbar}\), with the gas inlet regulated to maintain the target pressure at the \(0.5\%\) level~\cite{Ferraro2018LUNAGasTarget}. The windowless gas-target system developed for CASPAR also incorporates recovery, cleaning, and recompression of enriched target gas, although quantitative long-term pressure-stability data were not reported in the original system description~\cite{Robertson2016}. As a complementary example of differential-pumping performance under a high gas load, the Felsenkeller gas-jet target maintains a pressure of approximately \(4.5\times10^{-8}\,\mathrm{hPa}\) in the third differential stage over nozzle-inlet pressures up to \(6\,\mathrm{bar}\)~\cite{Yadav2026}. This value characterizes the isolation of the accelerator vacuum rather than the long-term stability of a recirculating target.

\begin{figure}[htbp]
    \centering
    \includegraphics[width=0.6\textwidth]{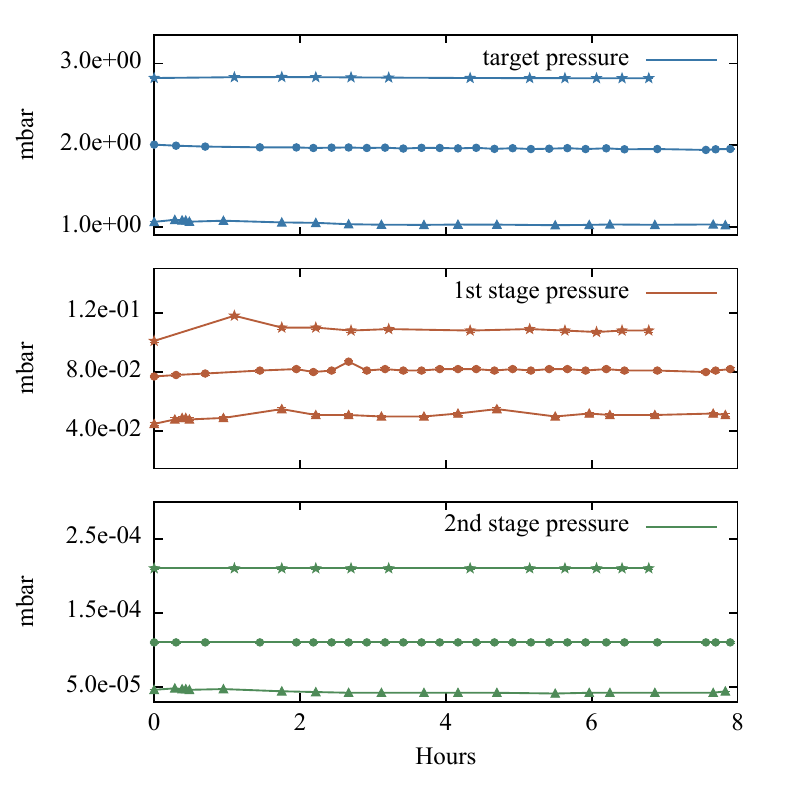}
    \caption{Pressure histories during eight-hour gas-circulation stability tests at target-chamber pressures of approximately 1, 2 and \(3\,\mathrm{mbar}\).}
    \label{fig:gas_circulation_test}
\end{figure}

To characterize the long-term stability of the JUNA system, closed-loop circulation tests were performed at target-chamber pressures of approximately 1, 2 and \(3\,\mathrm{mbar}\). Each operating condition was maintained continuously for \(8\,\mathrm{h}\), while the pressures in the target chamber and the first two differential stages were recorded simultaneously. Figure~\ref{fig:gas_circulation_test} shows the pressure histories. During stable operation, the target-chamber pressure fluctuated by less than \(1\%\).Throughout these measurements, the pressure in the third differential stage on the accelerator side remained at approximately \(10^{-4}\,\mathrm{Pa}\), demonstrating that the required high-vacuum condition could be maintained simultaneously with mbar-level target operation.

\begin{figure}[htbp]
    \centering
    \subfigure[]{\includegraphics[width=0.46\textwidth]{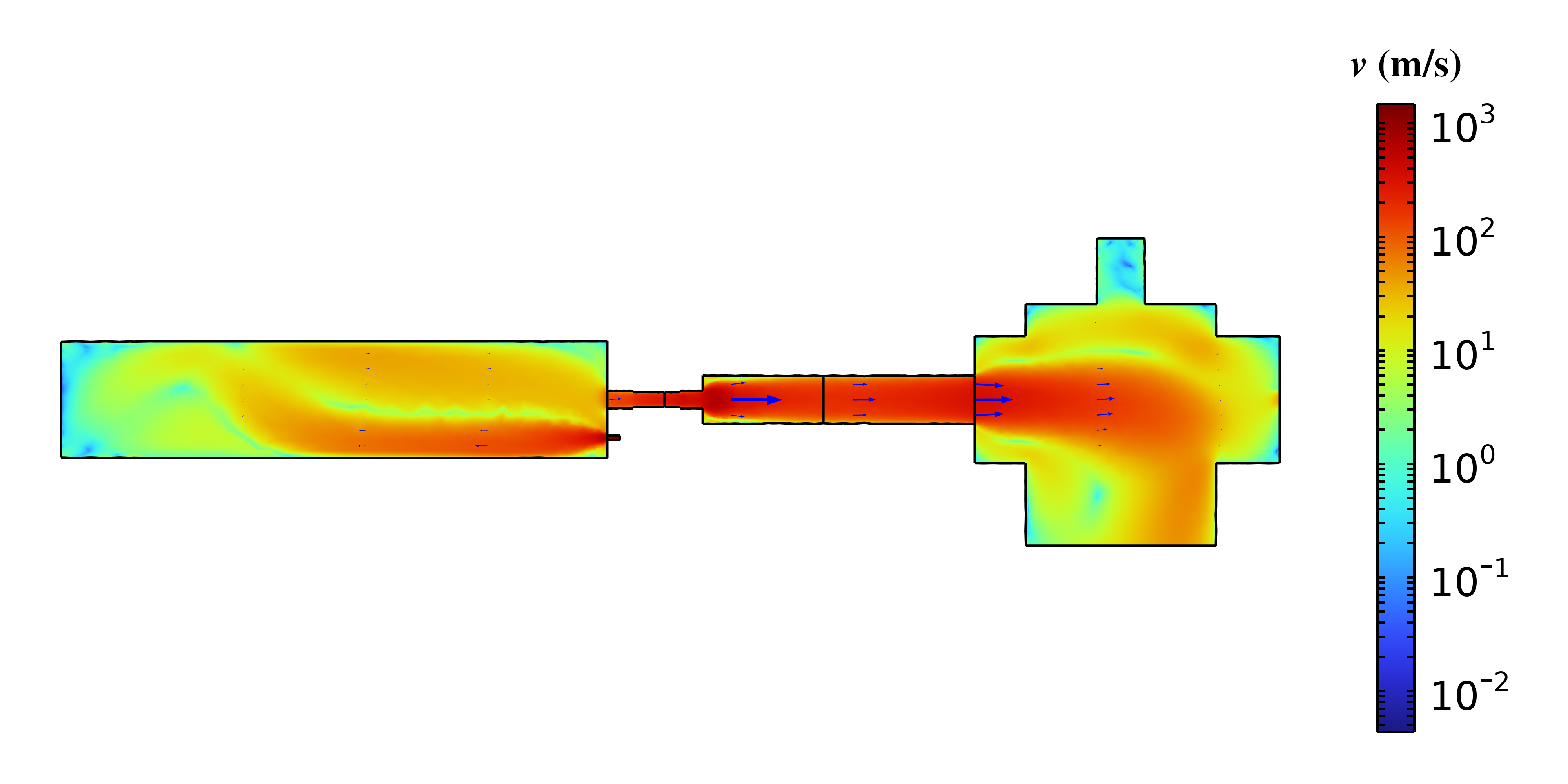}}%
    \hfill
    \subfigure[]{\includegraphics[width=0.46\textwidth]{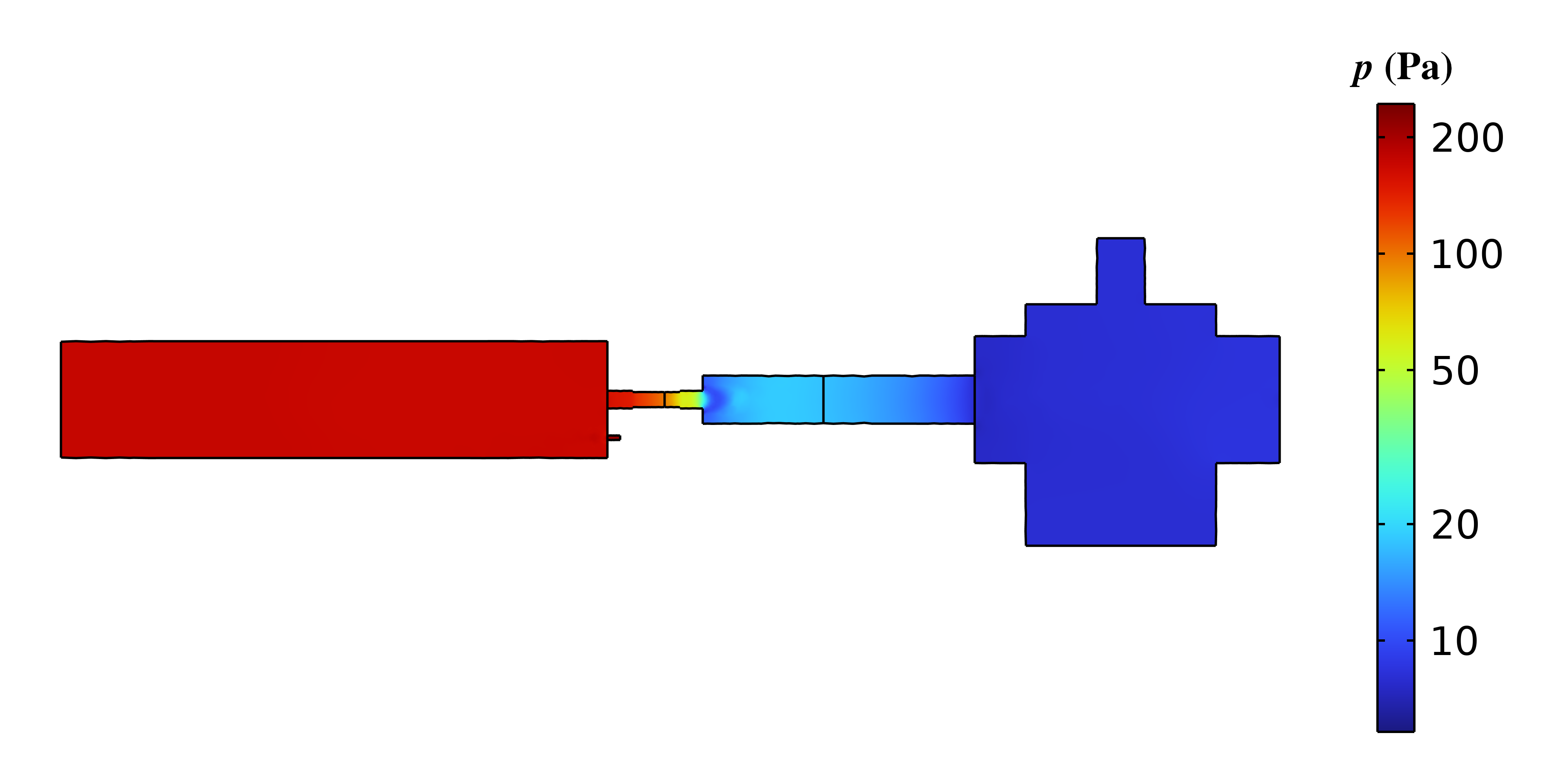}}
    \caption{Slip-flow simulation at \(1052\,\mathrm{sccm}\): (a) gas-speed field; (b) pressure field.}
    \label{fig:comsol_flow_fields}
\end{figure}

The gas-circulation flow rate was recorded simultaneously during the circulation tests. To investigate gas transport between the target chamber and the differential-pumping region under closed-loop operation, a numerical model including the target chamber, conductance-limiting tube, and first differential stage was established~\cite{Jiang2024WindowlessGasTarget}. As the pressure decreases from the mbar range in the target chamber to the Pa range in the first differential stage, the Slip Flow formulation was adopted to account for wall-slip effects in the low-pressure region. For a representative operating condition with the target chamber stabilized at \(2\,\mathrm{mbar}\), the measured circulation flow rate was \(1052\,\mathrm{sccm}\), and the calculated pressures in the target chamber and first differential stage agreed with the corresponding experimental values. The simulated velocity and pressure fields are shown in Fig.~\ref{fig:comsol_flow_fields}. Under this condition, the static pressure variation along the beam direction within the effective beam region remained below \(1.6\%\), which is of the same order of magnitude as the spatial pressure non-uniformity reported for the LUNA windowless gas target~\cite{Casella2002UndergroundSetup}.

\FloatBarrier

\subsection{Target Thickness Monitoring System}
Beam heating increases the gas temperature in the vicinity of the beam axis and consequently reduces the local gas number density~\cite{Gorres1980BeamHeating}. Under beam-on conditions, the gas density along the interaction region can therefore differ from that inferred from the static target-chamber pressure and temperature. As a result, the effective target thickness cannot in general be determined solely from the measured pressure, temperature, and geometrical length.

\begin{figure}[htbp]
    \centering
    \includegraphics[width=0.90\textwidth]{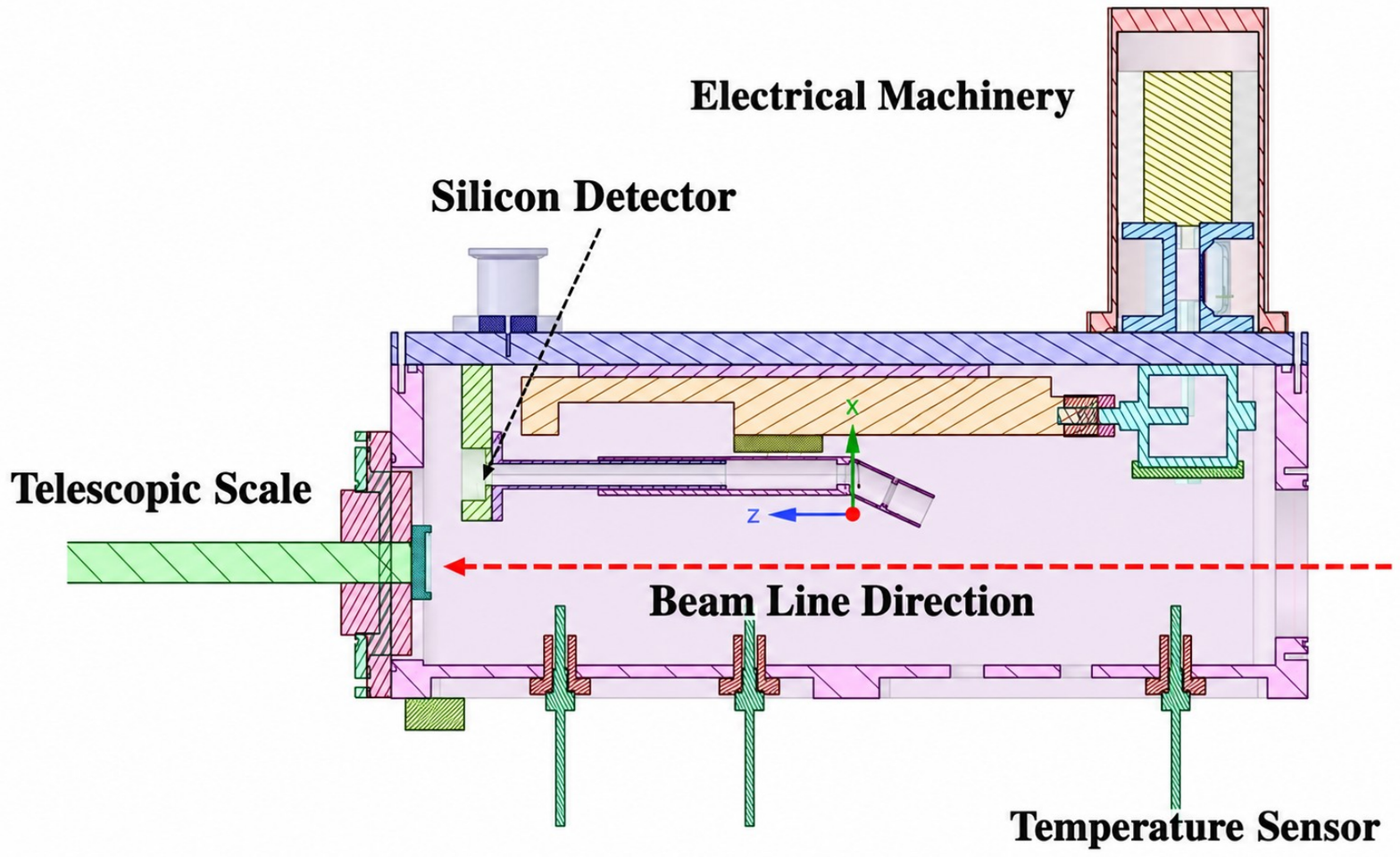}
    \caption{Geometry of the target-thickness monitoring system. The movable sampling tube scans along the beam axis, while secondary elastic scattering from the carbon foil provides additional geometrical and rate selection before the particles reach the silicon detector.}
    \label{fig:target_thickness_monitor}
\end{figure}

To obtain position-dependent information on the gas density during beam operation, a target-thickness monitoring system based on secondary elastic scattering was developed, as shown in Fig.~\ref{fig:target_thickness_monitor}. A movable sampling tube is installed at an angle of \(20^{\circ}\) with respect to the beam axis and can be translated along the interaction region by a motorized drive. A thin carbon foil is mounted at the bend of the tube. Taking the \(^{3}\mathrm{He}(\alpha,\gamma)^{7}\mathrm{Be}\) measurement as an example, incident \(\alpha\) particles are first elastically scattered by \(^{3}\mathrm{He}\) nuclei in the target gas. A fraction of the scattered particles enters the sampling tube and subsequently undergoes a second elastic-scattering process on the carbon foil before reaching the downstream silicon detector. Scanning the sampling tube along the beam axis therefore provides a position-dependent probe of the gas target under beam-on conditions.

The secondary-scattering stage is introduced primarily to reduce the particle rate incident on the silicon detector. According to the elastic-scattering measurements of Paneru et al.~\cite{Paneru2023SONIK}, the \(^{3}\mathrm{He}+{}^{4}\mathrm{He}\) cross section below \(E_{\alpha}=900\,\mathrm{keV}\) is largely described by Rutherford scattering. Using this approximation, the rate of primary elastically scattered \(\alpha\) particles at \(E_{\alpha}=900\,\mathrm{keV}\) and a beam current of \(2\,\mathrm{mA}\) is estimated to be \(R_{\rm pri}\approx9.6\times10^{8}\,\mathrm{s}^{-1}\). This corresponds to a mean interval of approximately \(1.0\,\mathrm{ns}\) between successive events, far below the characteristic processing time of conventional silicon-detector spectroscopy electronics.

For the geometry shown in Fig.~\ref{fig:target_thickness_monitor}, a \(15\,\mu\mathrm{g/cm^{2}}\) \(^{12}\mathrm{C}\) foil is used as the secondary-scattering foil, and \(^{4}\mathrm{He}\) particles scattered by approximately \(20^{\circ}\) are selected. The estimated counting rate at the silicon detector is thereby reduced to \(R_{\rm sec}\approx2.5\times10^{3}\,\mathrm{s}^{-1}\), corresponding to a transmission fraction of
\[
\frac{R_{\rm sec}}{R_{\rm pri}}
\approx
2.6\times10^{-6}.
\]
The resulting rate is within the practical operating range of conventional silicon charged-particle spectroscopy electronics, allowing the position-dependent scattering yield to be recorded without excessive detector count rate.






\FloatBarrier
\subsection{Calorimeter Structure and Simulation}

Charged particles undergo charge exchange while traversing the target gas. For the effective gas length of the present system, the charge-state distribution of the transmitted beam approaches equilibrium at a target pressure of approximately \(1\,\mathrm{mbar}\). Based on the measurements of Meckbach et al.~\cite{Meckbach1967}, an approximate estimate for an \(\alpha\) beam at \(E_{\mathrm{cm}}=380\,\mathrm{keV}\) indicates that the particle number inferred from the Faraday-cup charge is only about \(73\%\) of the actual value. Gas ionization can further affect the collected charge~\cite{Rudd1983ElectronCapture}, as can secondary-electron emission from beam-line components. We therefore developed a constant-temperature power-compensation calorimeter that is independent of the transmitted charge-state distribution.

\begin{figure}[htbp]
    \centering
    \subfigure[]{
        \includegraphics[width=0.47\textwidth]{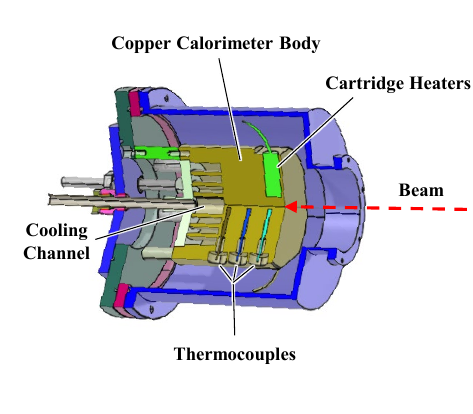}
        \label{fig:calorimeter_structure}}
    \hfill
    \subfigure[]{
        \includegraphics[width=0.47\textwidth]{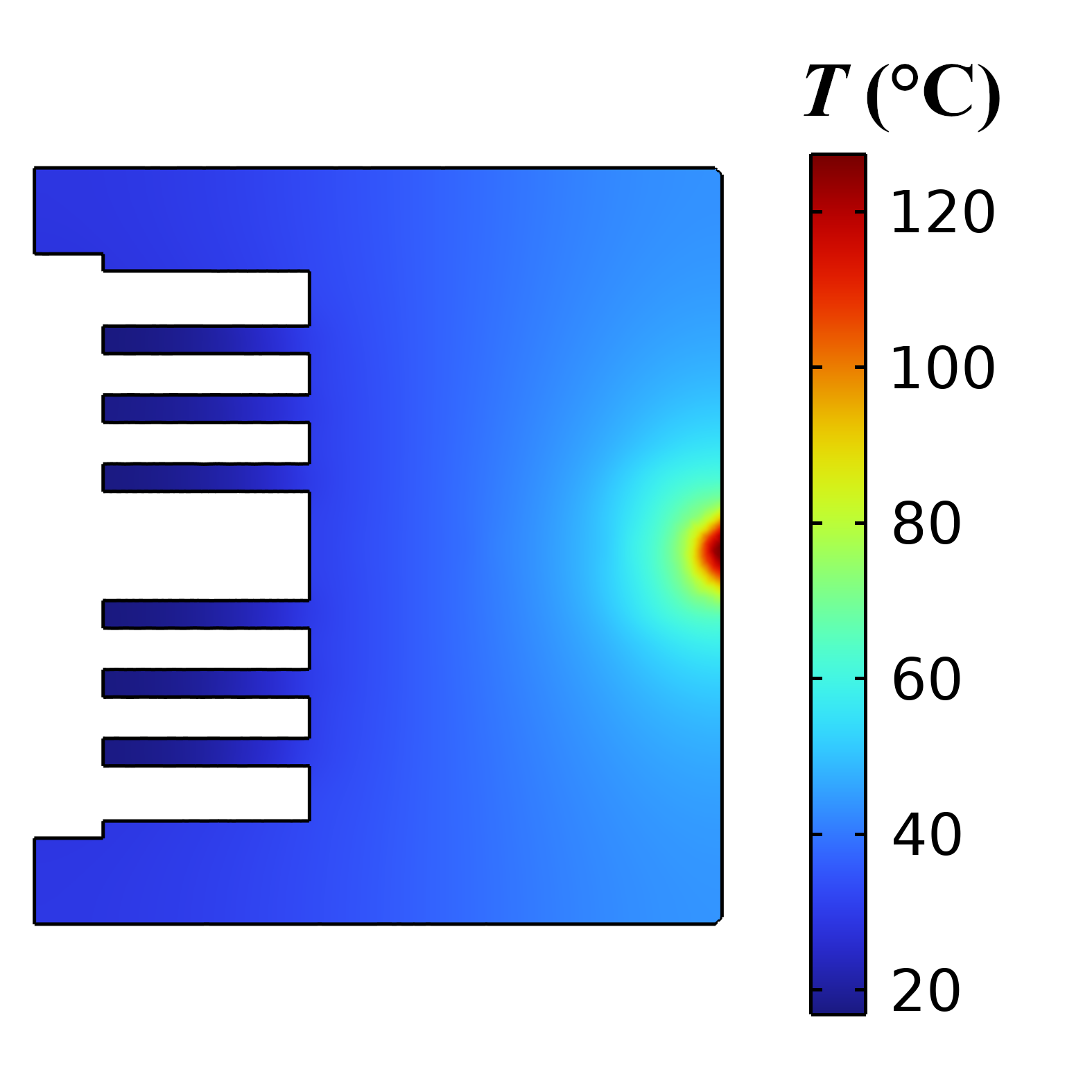}
        \label{fig:calorimeter_temperature_section}}
    \caption{Calorimeter structure and thermal simulation: (a) cutaway view of the calorimeter assembly; (b) simulated axial temperature distribution under the representative beam condition.}
    \label{fig:calorimeter_temperature}
\end{figure}

The calorimeter body is made of copper, with four cartridge heaters installed near the beam entrance to provide adjustable compensation power. Three thermocouples are arranged along the beam axis, and a water-cooling channel at the downstream end provides a stable heat-sink boundary. The overall structure and beam direction are shown in Fig.~\ref{fig:calorimeter_temperature}(a). The residual kinetic energy of particles reaching the calorimeter is deposited as heat.

The number of particles reaching the calorimeter can therefore be determined from the deposited thermal power as

\[
N_{\rm cal}=\int\frac{W_{\rm b}(t)}{E_{\rm cal}}\,\mathrm{d}t,
\]

where \(W_{\rm b}(t)\) is the deposited beam power and \(E_{\rm cal}\) is the energy deposited by each particle in the calorimeter.

The calorimeter operates in a constant-temperature power-compensation mode. Without beam, the cartridge-heater power is first set to \(W_0\). After the temperature at monitoring point \(i\) reaches a steady value, this temperature, \(T_{i,0}\), is taken as the reference. Following beam injection, the heater power is gradually reduced to \(W_{\rm h}\) until the monitoring point returns to and remains at \(T_{i,0}\). Because the beam and cartridge heaters differ in both source location and energy-deposition profile, equal changes in beam and heater power generally do not produce identical temperature responses at the same monitoring point. The thermal sensitivities of monitoring point \(i\) to heater power and deposited beam power are therefore defined as

\[
S_{{\rm h},i}=\frac{\Delta T_i}{\Delta W_{\rm h}},
\qquad
S_{{\rm b},i}=\frac{\Delta T_i}{\Delta W_{\rm b}}.
\]

For conditions relevant to the \(^{3}\mathrm{He}(\alpha,\gamma)^{7}\mathrm{Be}\) experiment, a centre-of-mass energy of \(E_{\mathrm{cm}}=380\,\mathrm{keV}\) and a \(^{4}\mathrm{He}^{2+}\) beam intensity of \(0.5\,\mathrm{pmA}\) correspond to an incident beam power of approximately \(443\,\mathrm{W}\). The simulated axial temperature distribution under this representative condition is shown in Fig.~\ref{fig:calorimeter_temperature}(b). All three thermocouples are located in smoothly varying regions of the temperature field, away from the steep gradient near the beam-deposition region, making them suitable reference points for constant-temperature power compensation.

\begin{figure}[!t]
    \centering
    \begin{minipage}[t]{0.72\textwidth}
        \centering
        \includegraphics[width=\linewidth]{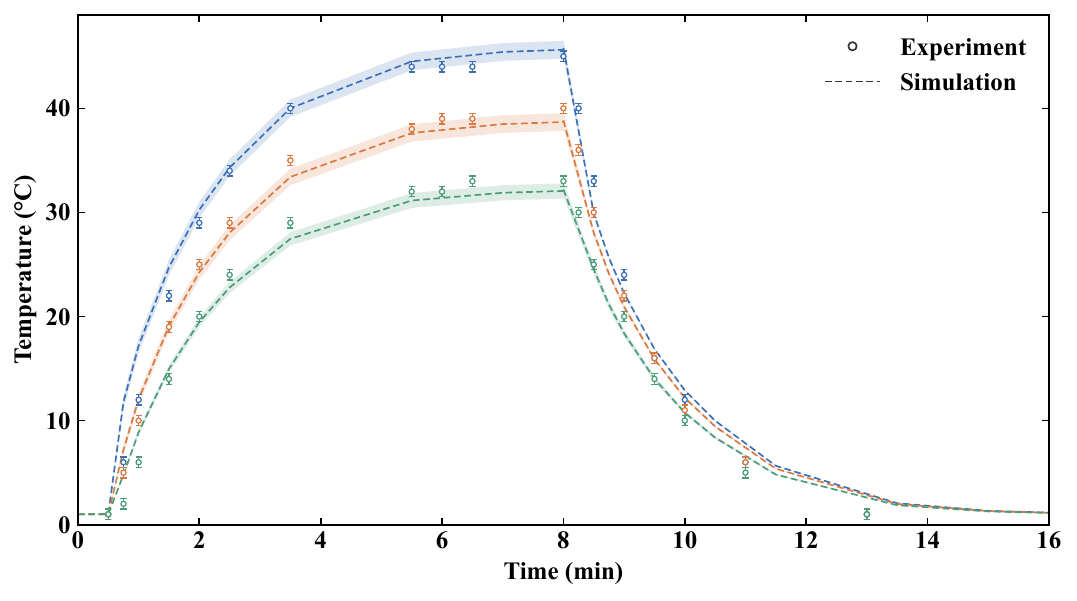}
        \par\smallskip (a)
    \end{minipage}
    \par\medskip
    \begin{minipage}[t]{0.41\textwidth}
        \centering
        \includegraphics[width=\linewidth]{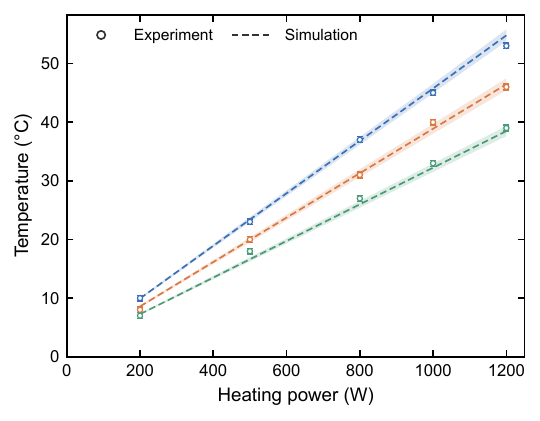}
        \par\smallskip (b)
    \end{minipage}
    \hfill
    \begin{minipage}[t]{0.43\textwidth}
        \centering
        \includegraphics[width=\linewidth]{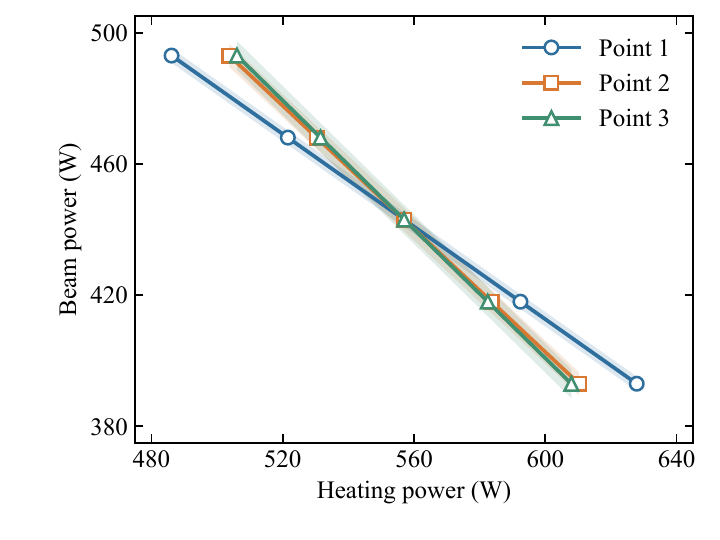}
        \par\smallskip (c)
    \end{minipage}
    \caption{
        Calorimeter response and beam-power conversion.
        (a) Transient temperature response;
        (b) steady-state temperature versus heater power;
        (c) beam power inferred using each monitoring point as the compensation reference.
        Symbols denote experimental measurements, while lines represent the thermal-model
        calculations. The shaded bands indicate the propagated power uncertainty associated
        with a \(0.1\,\mathrm{K}\) uncertainty in the temperature measurement.
    }
    \label{fig:calorimeter_sensitivity}
\end{figure}

The thermal model was evaluated under heater-only operation using both transient and steady-state measurements. Figure~\ref{fig:calorimeter_sensitivity}(a) shows the temperature evolution at the three monitoring points when a total heater power of \(1000\,\mathrm{W}\) was applied at \(t=30\,\mathrm{s}\) and switched off at \(t=8\,\mathrm{min}\). Figure~\ref{fig:calorimeter_sensitivity}(b) compares the measured temperatures with the corresponding model predictions over a cartridge-heater power range of \(200\text{--}1200\,\mathrm{W}\). The calculated temperature responses agree closely with the measurements over the investigated power range. A temperature-reading uncertainty of \(0.1\,\mathrm{K}\) was propagated through the local temperature--power relation to estimate the corresponding uncertainty in the inferred power. The model accurately reproduces not only the heating and cooling transients but also the absolute steady-state temperatures at all three monitoring points over the tested power range, demonstrating its capability to quantitatively describe both the transient and steady-state thermal responses of the calorimeter. Together with the approximately linear temperature--power response over the investigated operating range, these results support treating \(S_{{\rm h},i}\) and \(S_{{\rm b},i}\) as constants within their corresponding working ranges.

With unchanged cooling boundaries, the temperature changes produced by the reduction in heater power and by the deposited beam power balance when monitoring point \(i\) returns to its reference temperature. The deposited beam power is therefore given by

\[
W_{\rm b}
=
\left(W_0-W_{\rm h}\right)
\frac{S_{{\rm h},i}}{S_{{\rm b},i}}.
\]

Thus, the sensitivity ratio \(S_{{\rm h},i}/S_{{\rm b},i}\) provides the conversion between the change in compensation-heater power and the deposited beam power. The thermal model was used to determine the responses of the different monitoring positions to the two heat sources and thereby obtain the sensitivity correction required for constant-temperature power compensation.

Near the representative beam power of \(443\,\mathrm{W}\), the beam powers obtained from the sensitivity relation when each of the three monitoring points is used in turn as the constant-temperature reference are shown in Fig.~\ref{fig:calorimeter_sensitivity}(c). When monitoring point 2 is used as the compensation reference, the simulated sensitivities to beam power and heater power are \(S_{\rm b}=0.0404\,\mathrm{K/W}\) and \(S_{\rm h}=0.0379\,\mathrm{K/W}\), respectively, giving a sensitivity ratio of \(S_{\rm h}/S_{\rm b}\approx0.938\). Thus, directly equating the reduction in heater power with the deposited beam power would introduce a systematic deviation of approximately \(6.2\%\). The sensitivity correction arising from the different spatial distributions of the two heat sources must therefore be included in constant-temperature power-compensation measurements. These results demonstrate the necessity of the model-based power-conversion relation and provide the basis for quantitatively determining the particle flux reaching the calorimeter.

\FloatBarrier
\section{Beam commissioning measurements}

\subsection{Experimental setup and calibration}

Ground-level beam commissioning was performed using the \(600,\mathrm{kV}\) Cockcroft--Walton accelerator at the China Institute of Atomic Energy (CIAE) to characterize the operation of the windowless gas target under beam irradiation. The narrow resonance in \(^{14}\mathrm{N}(p,\gamma)^{15}\mathrm{O}\) and the direct-capture reaction \(^{12}\mathrm{C}(p,\gamma)^{13}\mathrm{N}\) were selected as benchmark reactions to examine the target response under different operating conditions.

Both measurements used the same windowless gas-target assembly. The proton beam passed through the three differential-pumping stages and the conductance-limiting tube before entering the target chamber. Reaction \(\gamma\) rays were detected by an external \(\mathrm{LaBr}_{3}(\mathrm{Ce})\) detector~\cite{VanLoef2002LaBr3}, and the transmitted beam was collected by a downstream Faraday cup. Target and differential-stage pressures, Faraday-cup current, live time, and \(\gamma\)-ray spectra were recorded for each run.

\begin{figure}[htbp]
    \centering
    \includegraphics[width=0.62\textwidth]{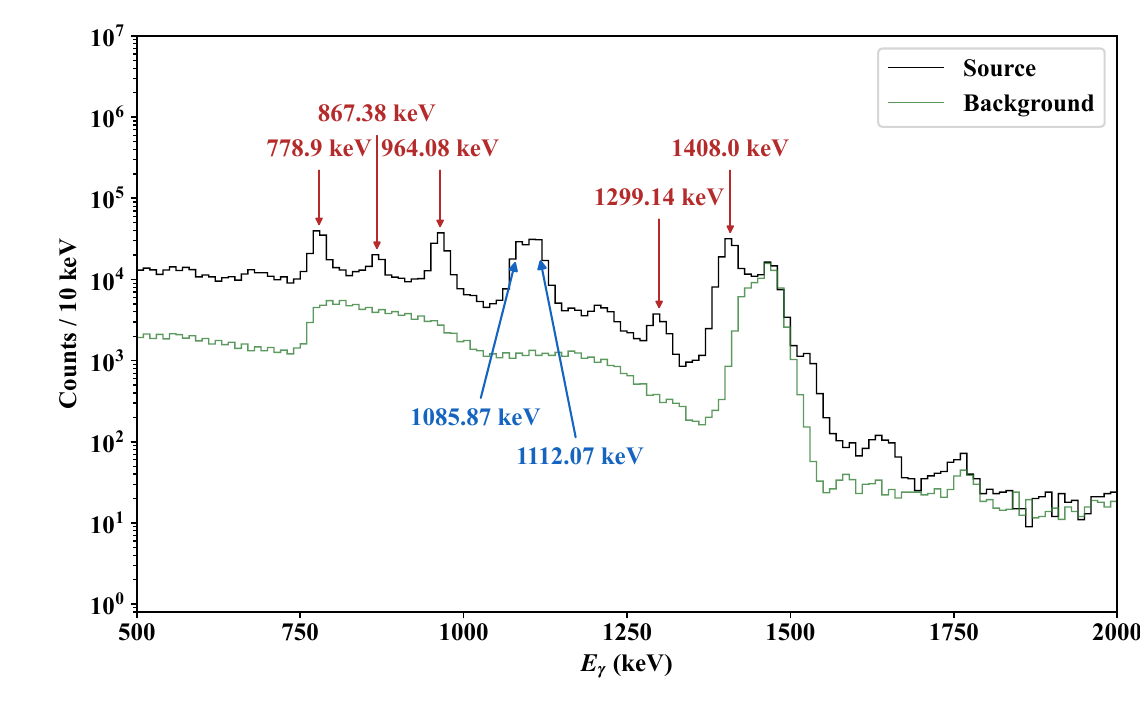}
    \caption{Energy-calibrated \(^{152}\mathrm{Eu}\) source spectrum (black) and live-time-normalized background spectrum (green). Peaks indicated by red arrows were included in the energy and full-energy peak-efficiency calibrations, whereas closely spaced peaks indicated by blue arrows were excluded because reliable peak separation was not possible.}
    \label{fig:eu152_spectrum}
\end{figure}

The detector energy response and full-energy peak efficiency were calibrated using a \(^{152}\mathrm{Eu}\) source. The source spectrum and the live-time-normalized background spectrum are shown in Fig.~\ref{fig:eu152_spectrum}. Peak centroids and net areas were obtained after background subtraction. The isolated peaks indicated by the red arrows were used for the energy and efficiency calibrations, whereas the closely spaced peaks indicated by the blue arrows were excluded because their areas could not be determined reliably. A linear relation between channel number and the known \(\gamma\)-ray energies was used for the energy calibration.

The source activity at the measurement time was calculated as
\[
A(t_{\mathrm{exp}})
=
A_{0}
\exp\left[
-\frac{t_{\mathrm{exp}}}{T_{1/2}}
\left(\ln 2\right)
\right],
\]
where \(A_{0}\) is the certified source activity and \(T_{1/2}\) is the half-life of
\(^{152}\mathrm{Eu}\). The absolute full-energy peak efficiency was then obtained from
\[
\varepsilon_E(E_{\gamma})
=
\frac{N_{\gamma}}
{t_{\mathrm{live}}A(t_{\mathrm{exp}})I_{\gamma}},
\]
where \(N_{\gamma}\), \(I_{\gamma}\), and \(t_{\mathrm{live}}\) are the net peak area,
emission probability, and live time, respectively. Its energy dependence was fitted using
\[
\ln \varepsilon_E(E_{\gamma})
=
\sum_{k=0}^{3} a_k
\left(\ln E_{\gamma}\right)^k.
\]

\begin{figure}[htbp]
    \centering
    \begin{minipage}[t]{0.47\textwidth}
        \centering
        \includegraphics[width=\linewidth]{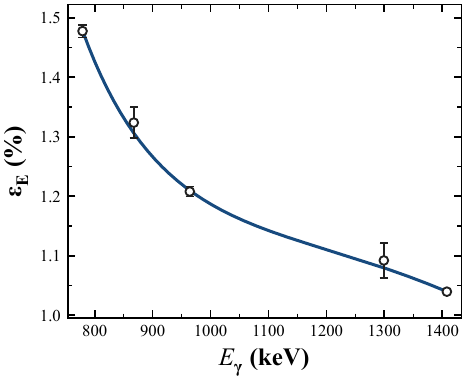}
        \par\smallskip (a)
    \end{minipage}
    \hfill
    \begin{minipage}[t]{0.47\textwidth}
        \centering
        \includegraphics[width=\linewidth]{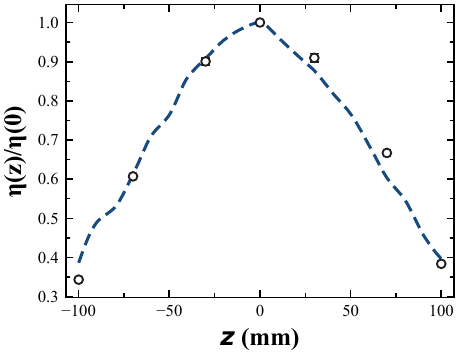}
        \par\smallskip (b)
    \end{minipage}
    \caption{\(\mathrm{LaBr}_{3}(\mathrm{Ce})\) detector efficiency calibration: (a) absolute full-energy peak efficiency as a function of \(\gamma\)-ray energy, with the curve representing the efficiency fit; (b) relative axial efficiency, with points representing the \(^{152}\mathrm{Eu}\) measurements and the dashed line representing the Geant4 calculation.}
    \label{fig:labr3_efficiency}
\end{figure}

The axial response was measured by placing the \(^{152}\mathrm{Eu}\) source at several positions along the beam axis. At each position, the efficiency relative to the detector-centre position was determined from the selected \(\gamma\)-ray lines. The detector geometry was also implemented in Geant4 to calculate the corresponding position dependence. As shown in Fig.~\ref{fig:labr3_efficiency}(b), the Geant4 calculation reproduces the measured axial variation over the investigated range. The position-dependent efficiency used in the subsequent analysis was written as
\begin{equation}
\varepsilon(E_{\gamma},z)
=
\varepsilon_E(E_{\gamma})f_z(z).
\label{eq:n14_efficiency}
\end{equation}
Here, \(f_z(z)\) is the relative axial response obtained from the Geant4 calculation and normalized at the detector-centre position.

The absolute full-energy peak efficiency and axial response are shown in Fig.~\ref{fig:labr3_efficiency}. In Fig.~\ref{fig:labr3_efficiency}(a), the points represent the \(^{152}\mathrm{Eu}\) measurements and the curve shows the fitted energy dependence. In Fig.~\ref{fig:labr3_efficiency}(b), the points represent the measured relative efficiencies and the dashed line shows the Geant4 calculation.

\FloatBarrier
\subsection{\(^{14}\mathrm{N}(p,\gamma)^{15}\mathrm{O}\) commissioning measurement}

The \(^{14}\mathrm{N}(p,\gamma)^{15}\mathrm{O}\) commissioning measurement was performed using \(\mathrm{N}_{2}\) as the target gas, with the incident proton energy scanned over \(300\text{--}310\,\mathrm{keV}\). After losing energy in the upstream beam line and target gas, the beam passed through the well-known narrow resonance at \(E_p=278\,\mathrm{keV}\), corresponding to the \(E_x=7556\,\mathrm{keV}\) excited state in \(^{15}\mathrm{O}\)~\cite{Formicola2004N14,Runkle2005N14,Bemmerer2006LUNA14N,Marta2010Resonances,Gyurky2019Activation}. The dominant decay branch of this state populates the \(6171.86\,\mathrm{keV}\) level with a branching ratio of \(58.4\pm0.3\%\)~\cite{Formicola2004N14}. The corresponding primary \(\gamma\) ray at approximately \(1384\,\mathrm{keV}\) was therefore selected for the yield analysis.

\begin{figure}[htbp]
    \centering
    \includegraphics[width=0.78\textwidth]{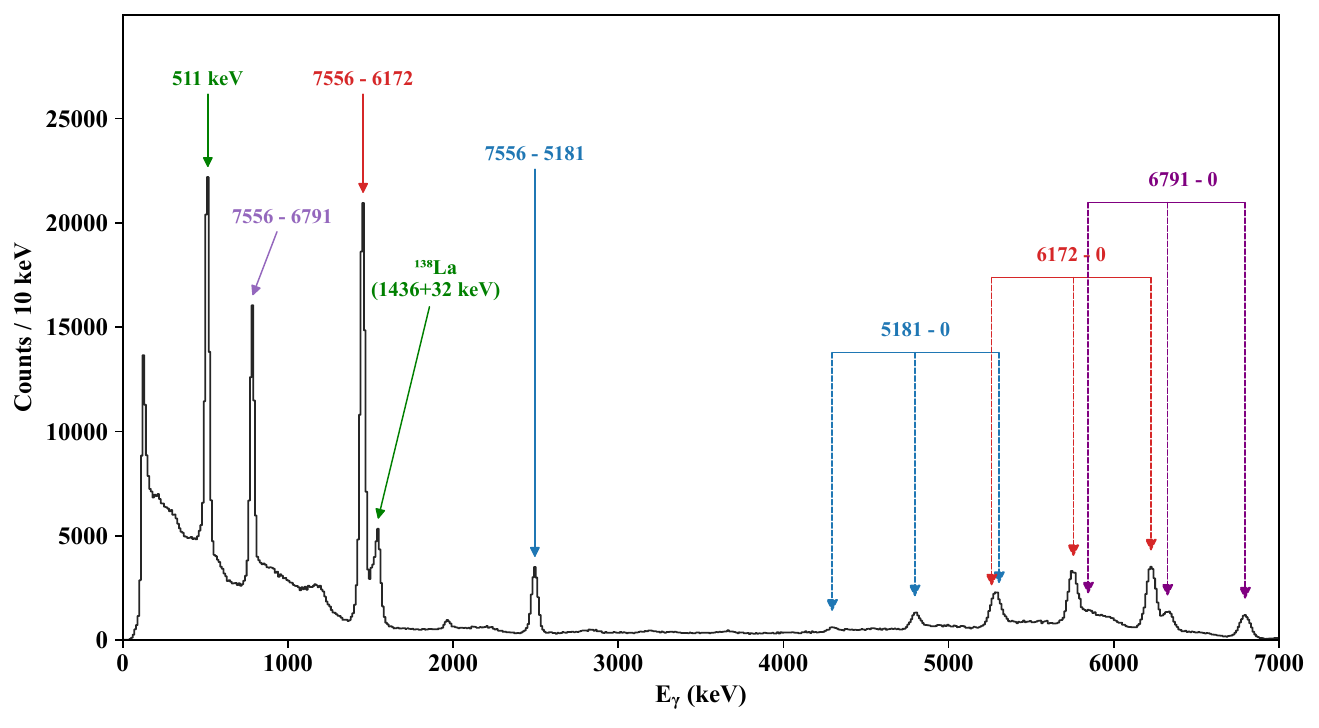}
    \caption{Representative \(\gamma\)-ray spectrum from the \(^{14}\mathrm{N}(p,\gamma)^{15}\mathrm{O}\) commissioning measurement over the full measured energy range. The principal primary and secondary transitions associated with the \(E_x=7556\,\mathrm{keV}\) resonance are indicated, together with the \(511\,\mathrm{keV}\) annihilation peak and the intrinsic \(^{138}\mathrm{La}\) background feature.}
    \label{fig:n14_spectrum}
\end{figure}

A representative \(\gamma\)-ray spectrum over the full measured energy range is shown in Fig.~\ref{fig:n14_spectrum}. The primary transitions from the \(7556\,\mathrm{keV}\) resonance state to the \(6791\), \(6172\), and \(5181\,\mathrm{keV}\) levels are identified, together with the subsequent transitions from these excited states to the ground state. The \(511\,\mathrm{keV}\) annihilation peak and the intrinsic \(^{138}\mathrm{La}\) background feature are also indicated in the spectrum. The simultaneous observation of the primary and secondary transitions provides a clear identification of the \(^{14}\mathrm{N}(p,\gamma)^{15}\mathrm{O}\) reaction and confirms the spectroscopic response of the \(\gamma\)-ray detection system during beam commissioning. Among these transitions, the strong \(7556\rightarrow6172\) primary transition near \(1384\,\mathrm{keV}\) was used for the subsequent yield analysis.

For each scan point, the net area of the \(1384\,\mathrm{keV}\) peak was extracted and normalized to the charge collected by the downstream Faraday cup. As the incident proton energy increased, the position at which the beam reached the narrow-resonance energy moved through the gas target towards the detector. Because the detection efficiency varies with the reaction position, the measured yield exhibited a corresponding energy dependence.

A Bayesian analysis was performed to describe the multi-energy scan data. The model incorporated the adopted resonance strength~\cite{Marta2010Resonances,Gyurky2019Activation}, effective stopping power, the \(1384\,\mathrm{keV}\) branching ratio~\cite{Formicola2004N14}, and the position-dependent detection efficiency shown in Fig.~\ref{fig:labr3_efficiency}(b). The beam-energy loss before the effective target region and an overall normalization factor were treated as free parameters and jointly constrained by the measured yields at all scan energies.

\begin{figure}[htbp]
    \centering
    \includegraphics[width=0.68\textwidth]{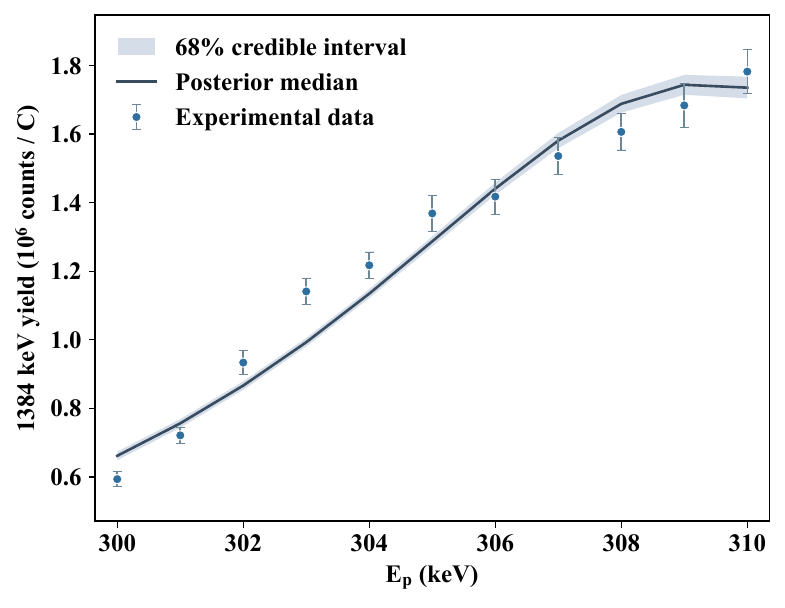}
    \caption{Bayesian analysis of the charge-normalized \(1384\,\mathrm{keV}\) yield as a function of incident proton energy. Points denote the measured data, the solid line represents the posterior median prediction, and the shaded band indicates the \(68\%\) posterior predictive interval.}
    \label{fig:n14_bayesian}
\end{figure}

Figure~\ref{fig:n14_bayesian} compares the charge-normalized measured yields with the posterior predictions. Over \(E_p=300\text{--}310\,\mathrm{keV}\), the resonance-reaction position moved through the target region towards the detector as the incident energy increased. The measured yield varied consistently with the detection efficiency at the corresponding reaction position, and the model reproduced this energy dependence well. These results show that the observed yield variation is consistent with the movement of the narrow-resonance reaction position within the target region.

\FloatBarrier
\subsection{\(^{12}\mathrm{C}(p,\gamma)^{13}\mathrm{N}\) commissioning measurement}

\begin{figure}[htbp]
    \centering
    \includegraphics[width=0.78\textwidth]{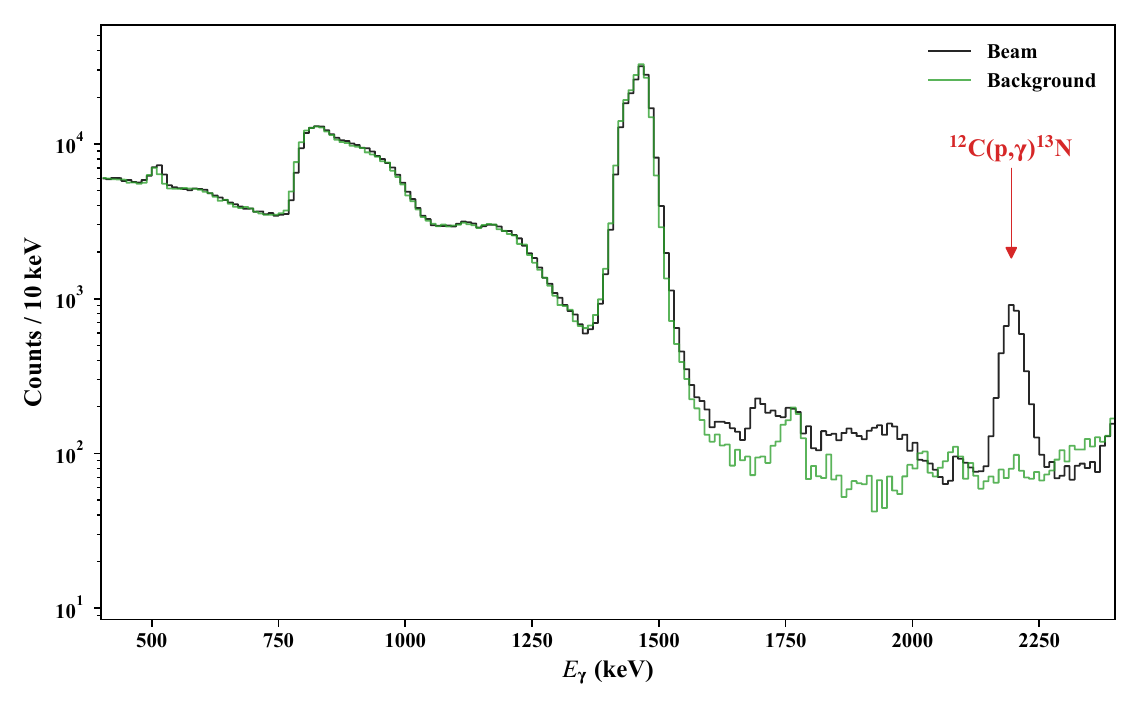}
    \caption{Measured \(^{12}\mathrm{C}(p,\gamma)^{13}\mathrm{N}\) spectrum at \(E_p=315\,\mathrm{keV}\). The ground-state capture peak used to determine the effective proton energy in the target is indicated.}
    \label{fig:12cpg_spectrum}
\end{figure}

Recent prompt-\(\gamma\) and activation measurements have provided high-precision data for the low-energy excitation function of \(^{12}\mathrm{C}(p,\gamma)^{13}\mathrm{N}\)~\cite{Gyurky2023C12pg,Skowronski2023C12pg,Csedreki2023C12pg,Kettner2023C12pg}, making this reaction a suitable benchmark for studying beam-heating effects in the gas target. Measurements were performed at \(E_p\approx315\,\mathrm{keV}\), and the corresponding \(\gamma\)-ray spectrum is shown in Fig.~\ref{fig:12cpg_spectrum}. The ground-state direct-capture peak of \(^{12}\mathrm{C}(p,\gamma)^{13}\mathrm{N}\) is clearly identified near \(2.2\,\mathrm{MeV}\).

Within the effective target region to which the detector has appreciable sensitivity, the proton energy loss along the beam direction is approximately \(9\,\mathrm{keV}\), corresponding to a change in the reaction cross section of less than \(9\%\). In comparison, the detection efficiency varies more strongly with the reaction position, as shown in Fig.~\ref{fig:labr3_efficiency}(b). The measured \(\gamma\)-ray yield is therefore predominantly weighted by the position-dependent detection efficiency, with reactions occurring closer to the detector contributing more strongly. This spatial weighting allows the centroid energy of the direct-capture peak to be used to infer the local proton energy in the effective reaction region near the detector.

For \(\gamma\) rays observed at approximately \(90^{\circ}\) with respect to the beam direction, the local proton energy in the laboratory frame, \(E_{p,\rm lab}\), is obtained by including the recoil of \(^{13}\mathrm{N}\):

\[
E_{p,\rm lab}
=
\frac{M_{13}}{M_{13}-m_p}
\left(
E_{\gamma}
+
\frac{E_{\gamma}^{2}}{2M_{13}c^{2}}
-
Q
\right),
\]

where \(m_p\) and \(M_{13}\) are the proton and \(^{13}\mathrm{N}\) masses,
respectively, and \(Q\) is the reaction \(Q\) value. The corresponding
effective beam-energy loss within the gas target was then determined as
\[
\Delta E_{\rm beam}
=
E_{p,\rm in}-E_{p,\rm lab},
\]
where \(E_{p,\rm in}\) denotes the proton energy at the entrance of the
effective gas-target region after accounting for the upstream beam-energy loss,
and \(E_{p,\rm lab}\) is the effective proton energy inferred from the measured
\(\gamma\)-ray peak centroid.

To quantify the change in effective gas density caused by beam heating, a
beam-heating factor was defined as

\[
h_{\rm beam}
=
\frac{\Delta E_{\rm beam}}
{\Delta E_{\rm theory}},
\]

where \(\Delta E_{\rm theory}\) is the proton energy loss calculated for the static target-gas density in the absence of beam heating. The molecular number density was obtained from the measured target pressure and temperature using the ideal-gas equation of state, and proton stopping powers in \(\mathrm{CO}_{2}\) were taken from SRIM~\cite{Ziegler2010SRIM} and interpolated to the corresponding proton energies. Since the stopping power varies only weakly over the relevant energy interval, \(h_{\rm beam}\) approximately represents the ratio of the effective beam-on gas density to the static gas density.

The extracted \(h_{\rm beam}\) values are shown in Fig. 14 as a function of the nominal beam-power deposition per unit target length estimated from the Faraday-cup current,

\[
\left(\frac{dW}{dx}\right)_{FC}
=
\frac{dE}{dx}\frac{I_{FC}}{e},
\]

where \(dE/dx\) is the proton stopping power in \(\mathrm{CO}_{2}\), \(I_{FC}\) is the current collected by the Faraday cup, and \(e\) is the elementary charge. The error bars were obtained by propagating the uncertainties of the fitted \(\gamma\)-ray peak centroids through the calculation of \(h_{\rm beam}\). The measured \(h_{\rm beam}\) decreases systematically with increasing deposited power per unit length, providing a direct characterization of the reduction in effective gas density induced by beam heating during gas-target operation.

\begin{figure}[htbp]
    \centering
    \includegraphics[width=0.66\textwidth]{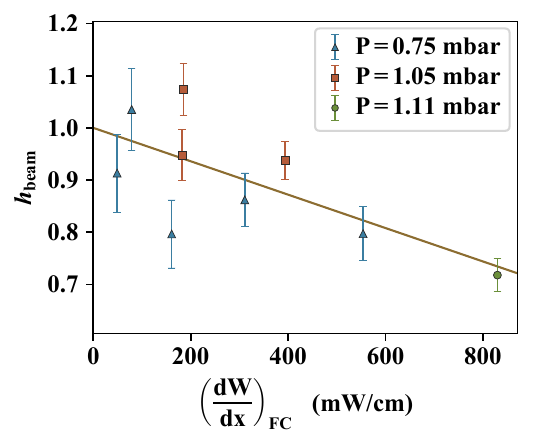}
    \caption{Effective beam-heating factor as a function of the nominal beam-power deposition per unit target length estimated from the Faraday-cup current, \((dW/dx)_{FC}\). Different symbols denote the target pressures, and the error bars include the propagated uncertainties of the fitted peak centroids.}
    \label{fig:12c_beam_heating}
\end{figure}


\FloatBarrier
\section{Implications for high-current gas-target operation}
\label{sec:commissioning_diagnostics}

The commissioning measurements provide direct information on the response of the gas target under beam irradiation. In the \(^{14}\mathrm{N}(p,\gamma)^{15}\mathrm{O}\) scans, the measured yield follows the change in detection efficiency as the narrow-resonance position moves through the target. The \(^{12}\mathrm{C}(p,\gamma)^{13}\mathrm{N}\) measurements similarly show a systematic reduction in effective gas density with increasing deposited beam power. These results establish the main trends associated with reaction position and beam heating under the tested operating conditions.

In both analyses, however, the beam intensity was derived from the charge collected by the downstream Faraday cup. For an extended gas target, charge exchange changes the charge-state distribution of the transmitted beam, while ionization of the target gas can introduce additional contributions to the collected charge. The Faraday-cup current therefore does not provide a sufficiently direct measure of the incident-particle number for an absolute comparison with other measurements. The commissioning results are consequently used here to characterize the relative response of the system rather than to establish an absolute beam-normalized benchmark.

The calorimeter and target-thickness monitor developed in this work extend these measurements toward a fully quantitative treatment. The calorimeter determines the beam intensity from deposited power without relying on the transmitted charge state, while the secondary elastic-scattering monitor provides position-dependent target information under beam irradiation. Together, they allow the beam normalization and effective target thickness to be determined independently, providing the additional constraints required for subsequent high-current measurements at JUNA.




\FloatBarrier
\section{Conclusions}

A high-current windowless gas-target system has been developed for low-energy nuclear-reaction measurements at JUNA. The system combines three-stage differential pumping, closed-loop gas recovery and purification, and dedicated diagnostics for beam intensity and effective target thickness. Stable circulation in the mbar pressure range was achieved while maintaining high vacuum on the accelerator side, demonstrating the capability required for extended operation with valuable isotopic gases. Gas-transport studies provided an estimate of the pressure variation within the target region, while thermal measurements and simulations established the sensitivity correction required for calorimetric beam determination. The position-resolved target-thickness monitor provides a dedicated approach for future beam-on characterization of the effective target density. Commissioning measurements with \(^{14}\mathrm{N}(p,\gamma)^{15}\mathrm{O}\) and \(^{12}\mathrm{C}(p,\gamma)^{13}\mathrm{N}\) demonstrated the combined operation of the gas target and \(\gamma\)-ray detection system and provided experimental information on the target response under beam conditions. Together, these results establish the principal operating and diagnostic capabilities of the JUNA windowless gas-target system.

The system is intended for subsequent measurements with isotopic gases, including \(^{3}\mathrm{He}(\alpha,\gamma)^{7}\mathrm{Be}\) and \(^{22}\mathrm{Ne}(\alpha,n)^{25}\mathrm{Mg}\). Efficient gas recovery and purification will reduce the consumption of \(^{3}\mathrm{He}\) and enriched \(^{22}\mathrm{Ne}\) and help maintain stable target conditions during extended measurements, while calorimetric beam normalization and position-resolved target monitoring provide the diagnostics required for precision measurements. Combined with the high-current beams and low-background underground environment of JUNA, these capabilities provide a basis for extending direct measurements of key nuclear-astrophysics reactions toward lower energies.



\section*{Acknowledgements}

This work was supported by the National Natural Science Foundation of China (Grant Nos. 12435010, 12605238, 11490560, 12475120, 12441507, 11875329, 12405162), the National Key R\&D Program of China (Grant No. 2022YFA1602301), the Jinping Deep Underground Frontier Science and Dark Matter Key Laboratory of Sichuan Province (Grant No. YLDC-ZBA-Z2025459), and the CNNC Fundamental Research Project.

\bibliography{mybibfile}

\end{document}